\documentclass[citeautoscript,floatfix,aps,prx,twocolumn,superscriptaddress,longbibliography]{revtex4-1}  %
\usepackage{graphicx}  %
\usepackage{dcolumn}   %
\usepackage{bm}        %
\usepackage{amssymb}   %
\usepackage{amsmath}
\usepackage{amsthm}
\newtheorem{theorem}{Theorem}
\newtheorem{lemma}[theorem]{Lemma}
\usepackage{braket}
\usepackage{amsfonts}
\usepackage{multirow}
\usepackage{dsfont}
\usepackage{hyperref}
\usepackage{xcolor}

\DeclareMathOperator{\arctanh}{arctanh}

\usepackage{mathtools}

\begin{document}

\title{Locality of gapped ground states in systems with power-law decaying interactions}
\author{Zhiyuan Wang}
\affiliation{Department of Physics and Astronomy, Rice University, Houston, Texas 77005,
  USA}
\affiliation{Rice Center for Quantum Materials, Rice University, Houston, Texas 77005, USA}
\author{Kaden R.~A. Hazzard}
\affiliation{Department of Physics and Astronomy, Rice University, Houston, Texas 77005,
USA}
\affiliation{Rice Center for Quantum Materials, Rice University, Houston, Texas 77005, USA}
\affiliation{Department of Physics and Astronomy, University of California, Davis, CA 95616, USA}
\date{\today}

\begin{abstract}
It has been proved that in gapped ground states of locally-interacting lattice quantum systems with a finite local Hilbert space, 
the effect of local perturbations decays exponentially with distance. %
However, in systems with power-law~($1/r^\alpha$) decaying interactions, no analogous statement has been shown, and there are serious mathematical obstacles to proving it with existing methods. 
In this paper we prove that when $\alpha$ exceeds the spatial dimension $D$,  
the effect of local perturbations on local properties a distance $r$ away is upper bounded by a power law $1/r^{\alpha_1}$ in gapped ground states,
provided that the perturbations do not close the spectral gap. The power-law exponent $\alpha_1$ is tight if $\alpha>2D$ and interactions are two-body, where we have $\alpha_1=\alpha$. The proof is enabled by a method that avoids the use of quasiadiabatic continuation and incorporates techniques of complex analysis. This method also improves bounds on ground state correlation decay, even in short-range interacting systems. 
Our work generalizes the fundamental notion that local perturbations have local effects to power-law interacting systems, 
with broad implications for numerical simulations and  experiments. 
\end{abstract}

\maketitle

\section{Introduction and overview of results}\label{sec:intro}
Locality is a fundamental principle that underlies many theories of nature. Loosely speaking, locality means that an object is influenced directly only by its immediate surroundings, and in particular,  should be insensitive to actions taken far away. %
The precise quantitative statement of this principle takes different forms in different contexts. In quantum many-body dynamics, locality manifests itself in the form of a causality lightcone: roughly, if a local perturbation takes place at time $t=0$, then at time $t$ its effect must be within a ball region $r\leq v t$, where $r$ is the distance and $v$ is the maximal allowed speed of propagation of any physical particles or signals in the system. In relativistic quantum field theories, such a causality lightcone is guaranteed by Lorentz invariance, where $v$ is the speed of light, and effects exactly vanish outside the lightcone. In non-relativistic quantum many-body systems with short-range interactions, the Lieb-Robinson bound~(LRB)~\cite{Lieb1972} guarantees an effective causality lightcone: the effect of local perturbations decays exponentially in $(r-vt)$, where the speed $v$ depends on the microscopic details of the system~\cite{hastings2006,schuch2011information,Wang2020Tightening}. 

Consequences of locality take a slightly different form for equilibrium properties of the  quantum many-body system. An important case is on the effect of a local perturbation on ground states. %
Specifically, let $\hat{H}$ be the Hamiltonian and consider the effect of a local perturbation $\hat{V}_Y$~(supported on region $Y$) on a local observable $\hat{S}_X$, supported on a region $X$ far from $Y$. Intuitively, we expect that the expectation value of $\langle \hat{S}_X\rangle $ measured in the perturbed ground state should not deviate significantly from its unperturbed value when the distance $d_{XY}$ is large, i.e. the deviation
\begin{equation}\label{def:delta_SX}
	\delta\langle \hat{S}_X \rangle_{\hat{V}_Y}\equiv \langle \hat{S}_X \rangle_{\hat{H}+\hat{V}_Y}-\langle \hat{S}_X \rangle_{\hat{H}}
\end{equation}
should be small in magnitude. 
This intuition is rigorously formulated as the  LPPL principle~(local perturbations perturb locally)~\cite{LPPL}, 
which states that for gapped ground states of a locally interacting Hamiltonian, $|\delta\langle \hat{S}_X \rangle_{\hat{V}_Y}|$ is upper bounded by a subexponentially decaying function in $d_{XY}$~\footnote{A function $f(x)$ decays subexponentially in $x$ if for any $\delta<1$, there
	exists an $x$-independent constant $c_\delta$ such that $ |f(x)|\leq
c_\delta e^{-|x|^\delta}$ at large $|x|$.}, provided that the perturbation does not close the spectral gap. The proof was based on the idea of quasiadiabatic continuation~(QAC)~\cite{LSMhigherD,hastings2005quasiadiabatic1,hastings2010locality}, which relates the perturbed ground state $|G\rangle_{\hat{H}+\hat{V}_Y}$ to the unperturbed one by a quasilocal unitary evolution 
\begin{equation}\label{eq:QAC}
	|G\rangle_{\hat{H}+\hat{V}_Y}=\mathcal{T}e^{i \int_0^1 H_{\mathrm{eff}}(t)dt} |G\rangle_{\hat{H}},
\end{equation}
where $\mathcal T$ is the time-ordering operation, and the effective Hamiltonian $H_{\mathrm{eff}}(t)$ only contains interactions that are subexponntially localized near $Y$. This immediately transforms the problem back to the dynamical case, where a Lieb-Robinson bound implies that $|\delta\langle \hat{S}_X \rangle_{\hat{V}_Y}|$ decays subexponentially in $d_{XY}$.
This bound %
was later strengthened to an exponential decay $|\delta\langle \hat{S}_X \rangle_{\hat{V}_Y}|\leq C e^{-\mu_1 d_{XY}}$~\cite{Roeck2015Local}\footnote{The method in Ref.~\cite{Roeck2015Local} does not use QAC. It is not clear to us if that method can be generalized to the power-law case, but when applied to short-range interacting systems, that method is more complicated than ours and gives looser bounds. Therefore we expect that even if that method can be generalized to the power-law case, the resulting bounds would be looser than our bounds listed in Tab.~\ref{tab:comp}. }, where $C$ is a constant and $\mu_1$ is given in Tab.~\ref{tab:comp}.  

In recent years, there has been increasing interest in %
understanding the analogous consequences of locality from 
long-range, power-law~($1/r^\alpha$) decaying interactions, driven in part by the ubiquity of these interactions  in many cold atom and molecule~\cite{Bloch2008Many,Aikawa2012Bose,yan2013observation,Seesselberg2018Extending,christakis2022probing}, Rydberg atom~\cite{bendkowsky2009observation,Saffman2010Quantum,bernien2017probing,Levine2019Parallel,Browaeys2020,Bluvstein2021Controlling,Bakr2021Quench}, and trapped ion~\cite{britton2012engineered,yao2012scalable,islam2013emergence,zhang2017observation,neyenhuis2017observation}  experiments, typically with $0\le \alpha\le 6$, as well as the Coulomb interaction. The important question then arises: when long-range interactions are present, to what extent can we still expect locality in the senses described above to hold? 
The answer to this question is far from obvious, since long-range interactions 
can give rise to non-local behaviors of correlation functions for sufficiently small $\alpha$~\cite{Eisert2013breakdown,Gorshkov2014}.
For the dynamical part, LRB has been successfully generalized to power-law  interacting systems~\cite{Foss2015nearly,tran2019locality,Lucas2019longrange,else2020improved,kuwahara2020strictly,Tran2020Hierarchy,Tran2021Lieb}, implying generalized causality lightcones~($r\propto e^{vt}$ for $D<\alpha<2D$~\cite{hastings2006}, $r= v t^\beta$ for $2D<\alpha<2D+1$~\cite{Foss2015nearly}, and $r= v t$ for $\alpha>2D+1$~\cite{Lucas2019longrange,kuwahara2020strictly}).  %

However, the implications of locality for equilibrium systems are far less understood when power-law interactions are present, even in the important case of gapped ground states. 
This is partly due to the difficulties caused by the appearance of long-range interactions in $H_{\mathrm{eff}}(t)$ in Eq.~\eqref{eq:QAC}: %
QAC only leads to a LPPL bound for $\alpha > 2D$~\cite{Gong2017Entanglement}, an extremely restrictive condition and one rarely satisfied in the experimental systems of interest. 
Furthermore, even for $\alpha>2D$, the LPPL principle has never been proved, and the above method with the QAC in Ref.~\cite{Gong2017Entanglement} would lead to power law exponents in the resulting bounds that are not tight~(see App.~\ref{appen:LPPL_QAC} for details). 

In this paper, we prove the LPPL principle for gapped ground states of  
lattice quantum systems where interactions are bounded by a power law $1/r^\alpha$ in distance $r$, with $\alpha>D$. To achieve this goal, we devise an alternative method that avoids the use of QAC Eq.~\eqref{eq:QAC}~(thereby circumventing the aforementioned difficulty) and incorporates techniques of complex analysis. This method also improves the LPPL bounds for short-range interacting systems, and applies to degenerate~(either exact or approximate) ground states as well.  
Our main result is roughly as follows: for perturbations $\hat{V}_Y$ that do not close the spectral gap,
\begin{equation}\label{eq:main_result}
	|\delta\langle \hat{S}_X \rangle_{\hat{V}_Y}| \leq \begin{cases}
		\mathbf{P}(\ln d_{XY})/d_{XY}^{\alpha_1}\\
		\mathbf{P}(d_{XY}) e^{-\mu_1 d_{XY}},
	\end{cases}%
\end{equation}
where  $\langle \ldots \rangle$ is a uniform average %
over the~(possibly degenerate) ground state subspace, the first line is for power-law systems and the second line is for short-range interacting systems, the exponents $\alpha_1,\mu_1$ are given in Tab.~\ref{tab:comp}, and throughout this paper we use $\mathbf{P}(x)$ to denote a polynomial in $x$ with non-negative coefficients~[but $\mathbf{P}(x)$ in different equations or in different parts of the same equation need not be the same]~\footnote{The $\mathbf{P}(x)$ in Eqs.~(\ref{eq:main_result},\ref{eq:correlation_decay_bound}) is at most quadratic in $x$, while the $\mathbf{P}(x)$ in Eq.~\eqref{eq:FSEB} is at most cubic in $x$.}. We see that  $\alpha_1$  is equal to $\alpha$ if $\alpha>2D$ and interactions are two-body, in which case our bound is qualitatively tight~[up to the subleading prefactor $\mathbf{P}(\ln d_{XY})$] since it agrees with perturbation theory. 

As one notable byproduct, the method we use to obtain these bounds also improves bounds on  correlation decay~\cite{hastings2006,nachtergaele2006} of gapped~(possibly degenerate) ground states: for arbitrary local operators $\hat{A}_X,\hat{B}_Y$, their  connected correlation function   is bounded by %
\begin{equation}\label{eq:correlation_decay_bound}
	 |\langle \hat{A}_X \hat{B}_Y\rangle-\langle \hat{A}_X \rangle\langle  \hat{B}_Y\rangle|\leq \begin{cases}
		\mathbf{P}(\ln d_{XY})/d_{XY}^{\alpha_2}\\
		\mathbf{P}(d_{XY}) e^{-\mu_2 d_{XY}},
	\end{cases}%
\end{equation}
where the exponents $\alpha_2,\mu_2$ are given in Tab.~\ref{tab:comp}. We see that our method improves earlier exponents, even in the case of short-range interacting systems, where our bound improves Ref.~\cite{hastings2006}'s bound by approximately a factor of 2 for $\Delta\ll v$. %

Our results have profound implications on numerical simulations and  experiments. For example, it has been pointed out~\cite{Wang2021Bounding} that the LPPL principle straightforwardly implies an upper bound on the finite size error of several numerical ground state algorithms, such as exact diagonalization~\cite{noack:diagonalization_2005,
	sandvik2010computational} and the density matrix renormalization group~\cite{white:density_1992,schollwoeck:density_2011}. Our results Eq.~\eqref{eq:main_result} imply that 
the finite size error of a local observable $\hat{S}$ in gapped ground state simulations decays in the linear dimension of the system $L$ as
\begin{equation}\label{eq:FSEB}
	\delta\langle\hat{S}\rangle_L\equiv |\langle\hat{S}\rangle_L-\langle\hat{S}\rangle_\infty|\leq \begin{cases}
		\mathbf{P}(\ln L)/L^{\alpha_3}\\
		\mathbf{P}(L) e^{-\mu_3 L},
	\end{cases}%
\end{equation}
provided that the finite system is connected to the thermodynamic limit by a uniformly gapped path~\cite{Wang2021Bounding}. As in Eqs.~(\ref{eq:main_result},\ref{eq:correlation_decay_bound}) the first line is for power-law systems while the second line is for short-range interacting systems, and
the constants  $\alpha_3,\mu_3$ are given in Tab.~\ref{tab:comp}. 

Our paper is organized as follows. Tab.~\ref{tab:comp} summarizes the exponents $\alpha_1,\alpha_2,\alpha_3,\mu_1,\mu_2,\mu_3$ in Eqs.~(\ref{eq:main_result},\ref{eq:correlation_decay_bound},\ref{eq:FSEB}) for various interaction ranges. In Sec.~\ref{sec:main} we introduce our improved method, and use this method to bound the response of local observables in gapped non-degenerate ground states,  and obtain the main result, Eq.~\eqref{eq:main_result}. In Sec.~\ref{sec:degenerate} we generalize the bounds to gapped degenerate ground states. In Sec.~\ref{sec:FSE} we discuss the implications of our bounds in finite size numerical simulations and prove Eq.~\eqref{eq:FSEB}. In Sec.~\ref{sec:correlation} we use our improved method to obtain tighter bounds on ground state correlation decay, Eq.~\eqref{eq:correlation_decay_bound}. We conclude in Sec.~\ref{sec:conclusions}.

\begin{widetext}
\begin{center}
 \begin{table}[ht]
	\centering
	{\renewcommand{\arraystretch}{1.5}
		\begin{tabular}{|c|c|c|c|c|}
			\hline
			\multirow{2}{*}{Interaction}   & \multicolumn{2}{c|}{Prior bound}  & Our bound~(LPPL and correlation decay & FSE bound \\
			\cline{2-3}
			                                     & LPPL  &  Correlation decay  &    have  same exponents:  $\alpha_1=\alpha_2,~\mu_1=\mu_2$)      &          \\
			\hline
			\begin{tabular}{@{}c@{}} $1/r^\alpha$, \\$ \alpha>D$\\ 
			\end{tabular}
			 &       -         &    $\alpha_2=\frac{\alpha}{1+2v/\Delta}$~\cite{hastings2006} &$\alpha_1=\frac{2\alpha}{\pi}\arcsin(\tanh\frac{\Delta\pi}{2v})$   & \begin{tabular}{@{}c@{}}if $\alpha>D+1$:\\$\alpha_3=\min(\alpha-D,\alpha_1+1-D)$\\ if $D<\alpha\leq D+1$: \\ $\alpha_3=\begin{cases}
					\alpha-D   &\text{\!\!\!if } \alpha_1>D\\
					\alpha_1+\alpha-2D & \text{\!\!\!if } \alpha_1\leq D
				\end{cases}$\end{tabular}
			   \\
			\hline 
			\begin{tabular}{@{}c@{}} $1/r^\alpha, \alpha> 2D $ \\ two body \\ 
			\end{tabular}
			 &        -        & $\alpha_2=\alpha$~\cite{Tran2020Hierarchy}  &    $\alpha_1=\alpha$  &$\alpha_3=\alpha-D$\\
			\hline
			$e^{-\mu r}$ &    $\mu_1=\frac{\mu}{1+2\mu v/\Delta}$~\cite{Roeck2015Local}      &   $\mu_2=\frac{\mu}{1+2\mu v/\Delta}$~\cite{hastings2006,Hastings2006Note}   & $\mu_1=\frac{2\mu}{\pi}\arcsin(\tanh\frac{\Delta\pi}{2\mu v})$
		    & $\mu_3=\mu_1$\\
			\hline
	\end{tabular}}
	\caption{\label{tab:comp} Summary of the constants $\alpha_1$, $\mu_1$~(LPPL bounds), $\alpha_2$, $\mu_2$~(correlation decay bounds), and $\alpha_3$, $\mu_3$~(finite size error bounds) for previous results compared with ours, 
		for both power-law and short-range interacting systems. 
		Our main result is the proof of the LPPL principle Eq.~\eqref{eq:main_result} for ground states of power-law interacting systems with spectral gap $\Delta$, but we also significantly improved the  bound for   systems with exponentially-decaying interactions, as well as the constants $\alpha_2,\mu_2$ that appears in the correlation decay bounds Eq.~\eqref{eq:correlation_decay_bound}. The FSE bound Eq.~\eqref{eq:FSEB} with exponents $\alpha_3,\mu_3$ is a primary application of our main result~[previously, there is only a FSE bound for short-range systems~\cite{Wang2021Bounding}, in which $\mu_3=\mu_1=\mu/(1+2\mu v/\Delta)$].  $v$ is a constant that appears in the LRB that can be straightforwardly calculated~(for short-range interacting systems, $v$ is the Lieb-Robinson speed).}
\end{table}
\end{center}
\end{widetext}

\section{Locality of perturbations to  gapped non-degenerate ground states}\label{sec:main}
Our  set-up is as follows. Let $\Lambda_L$ be an infinite sequence of $D$-dimensional finite lattices, labeled by the linear system size $L\in\mathbb{Z}$, with $N\propto L^D$ number of lattice sites in total. On each site $i\in \Lambda_L$ sits a quantum degree of freedom with local Hilbert space $\mathcal{H}_i$. In this paper we focus on fermionic systems or quantum spin systems where $\mathcal{H}_i$ is finite dimensional, although our formalism can be straightforwardly generalized to bosonic systems where $\mathrm{dim}(\mathcal{H}_i)$ is infinite. The Hamiltonian $H_L$ acts on the global Hilbert space $\mathcal{H}_L\equiv\bigotimes_{i\in \Lambda_L} \mathcal{H}_i$, and can be written in the generic form 
\begin{equation}\label{eq:Hamiltonian}
	H_L=\sum_{X\subset \Lambda_L} h_X,
\end{equation}
where the summation is over all subsets of $\Lambda_L$ and $h_X$ is the local Hamiltonian supported on $X$~\footnote{For fermionic systems, $h_X$ is an operator that is even in the fermion creation/annihilation operators, and only involve fermionic modes inside the region $X$. This is enough to guarantee the important condition for the LRBs used later in this paper: for non-overlapping regions $X_1,X_2$~(i.e. $X_1\cap X_2=\emptyset$), we always have $[h_{X_1},h_{X_2}]=0$.}~(we will later specify some locality condition on $h_X$ which requires $\|h_X\|$ to be small for large $X$). Throughout this section we assume that $H_L$ has a non-degenerate ground state $|G_L\rangle$ with spectral gap $\Delta_L$~(the energy difference between the first excited state and the ground state) that is uniformly bounded from below, i.e. there exists $\Delta^{(0)}>0$ such that $\Delta_L\geq \Delta^{(0)}$ for all $\Lambda_L$. %
At this point we do not make assumptions on the range of interaction, nor do we assume that the local Hilbert space is finite dimensional. 

Let $V_Y$ be a local perturbation supported on region $Y$. Suppose that for all $\lambda\in[0,1]$, $H_L(\lambda)\equiv H_L+\lambda V_Y$ has a non-degenerate ground state $|G_L(\lambda)\rangle$ with spectral gap $\Delta_L(\lambda)$ that is uniformly bounded from below, i.e. $\exists \Delta>0$ such that $\forall \lambda\in[0,1]$, $\Delta_L(\lambda)\geq \Delta>0$, for all $\Lambda_L$. 
This condition will always be satisfied for sufficiently small perturbations satisfying $||V_Y||<\Delta^{(0)}/2$~($||\cdots||$ is the operator norm), since Weyl's inequality~\cite{bhatia1996matrix} gives $\Delta_L(\lambda)\ge \Delta_L - 2\lambda ||V_Y|| \ge \Delta^{(0)} - 2||V_Y||$.

Let $S_X$ be a local observable supported on region $X$ such that $X\cap Y=\emptyset$. Our goal is to bound the response of $S_X$ to the local perturbation $V_Y$, as defined in Eq.~\eqref{def:delta_SX}. We achieve this goal in two steps: in Sec.~\ref{sec:improved_method} we present a general method to bound $\delta\langle \hat{S}_X \rangle_{\hat{V}_Y}$ using a Lieb-Robinson-type bound on  the unequal time correlator $\langle G_L(\lambda)| [S_X(t), V_Y]|G_L(\lambda)\rangle$, where $S_X(t)=e^{iH_Lt} S_X e^{-iH_Lt}$, and then in Secs.~\ref{sec:LPPL_large_alpha}-\ref{sec:LPPL_local} we specialize to systems with different interaction ranges and apply the corresponding Lieb-Robinson bounds %
to obtain our main results in Eq.~\eqref{eq:main_result} and Tab.~\ref{tab:comp}. The resulting bounds are independent of the system size $L$, so they hold in the thermodynamic limit $L\to\infty$.

\subsection{The improved method}\label{sec:improved_method}
In the following we present an improved method to bound $\delta\langle \hat{S}_X \rangle_{\hat{V}_Y}$ using a Lieb-Robinson-type bound on $\langle G_L(\lambda)| [S_X(t), V_Y]|G_L(\lambda)\rangle$. %
There are two main improvements compared to previous approaches: the first part generalizes the method in Ref.~\cite{Wang2021Bounding}, which avoids the QAC and directly relates $\delta\langle \hat{S}_X \rangle_{\hat{V}_Y}$ to a specially constructed correlation function, while the second part obtains a bound on  this correlation function from a LRB on $|\langle G_L(\lambda)| [S_X(t), V_Y]|G_L(\lambda)\rangle|$ using complex analysis techniques, which significantly improves the previous method in Ref.~\cite{Wang2021Bounding}. %

Since we have a gapped path for $\lambda\in[0,1]$,  we can use perturbation theory to relate the rate of change of $\langle \hat{S}_X\rangle_{L,\lambda}\equiv \langle G_L(\lambda)| \hat{S}_X|G_L(\lambda)\rangle$ at each $\lambda$ to a special correlation function, from which we will obtain an exact expression for $\delta\langle \hat{S}_X \rangle_{\hat{V}_Y}$ as an integral over the  correlation function. Choose the normalization and phase of  $|G_L(\lambda)\rangle$ such that $\langle G_L(\lambda)|G_L(\lambda)\rangle=1$ and $\langle G_L(\lambda)|\frac{d}{d\lambda}|G_L(\lambda)\rangle=0, \forall\lambda\in [0,1]$. For any finite $L$, first order non-degenerate perturbation theory gives the exact identity
\begin{equation}\label{eq:PTevolution}
	\frac{d}{d\lambda}|G_L(\lambda)\rangle=\frac{\bar{P}_{G_L}(\lambda)}{\hat{H}_L(\lambda)-E_L(\lambda)}V_Y|G_L(\lambda)\rangle,
\end{equation}
where $E_L(\lambda)$ is the ground state energy of $\hat{H}_L(\lambda)$ and $\bar{P}_{G_L}(\lambda)\equiv \hat{\mathds{1}}-|G_L(\lambda)\rangle\langle G_L(\lambda)|$ is the projection operator to the space of excited states. Then %
\begin{equation}\label{eq:deltaSX}
	\frac{d}{d\lambda}\langle \hat{S}_X \rangle_{L,\lambda}=\langle G_L(\lambda)|\hat{S}_X\frac{\bar{P}_{G_L}(\lambda)}{\hat{\Delta}_L(\lambda)}\hat{V}_Y|G_L(\lambda)\rangle+\mathrm{c.c.},
\end{equation}
where $\hat{\Delta}_L(\lambda)\equiv \hat{H}_L(\lambda)-E_L(\lambda)$, whose spectrum is lower bounded by $\Delta$. In the following we prove a uniform bound~(independent of $L,\lambda$) on the RHS of Eq.~\eqref{eq:deltaSX}, so that a bound on $\delta\langle \hat{S}_X \rangle_{\hat{V}_Y}$ immediately follows from $	|\delta\langle \hat{S}_X \rangle_{\hat{V}_Y}|\leq \int_0^1 d\lambda |d\langle \hat{S}_X \rangle_{L,\lambda}/d\lambda| $.

From now on we omit the labels $L,\lambda$. %
We define
\begin{equation}\label{eq:G_XY_spectral_decomp}
	\Omega_{XY}(\omega)\equiv\langle G|\hat{S}_X\frac{i\bar{P}_G}{\omega-\hat{\Delta}}\hat{V}_Y|G\rangle-\langle G|\hat{V}_Y\frac{i\bar{P}_G}{\omega+\hat{\Delta}}\hat{S}_X|G\rangle
\end{equation}
in the region $\omega\in \mathbb{C}\backslash K_\Delta$, where $K_\Delta=\{\omega\in\mathbb{R}|\omega\geq \Delta\text{ or }\omega\leq -\Delta\}$.  %
Notice that for any finite system size $L$, $\Omega_{XY}(\omega)$ is a complex analytic function in its domain. Furthermore, the RHS of Eq.~\eqref{eq:deltaSX} is exactly $i \Omega_{XY}(0)$. For $|\mathrm{Im}(\omega)|> 0$, we have an integral representation for $\Omega_{XY}(\omega)$
\begin{eqnarray}\label{eq:G_XY_integral_rep}
	\Omega_{XY}(\omega)=\int^{\eta_\omega\infty}_0\langle G|[\hat{S}_X(t),\hat{V}_Y]|G\rangle e^{i\omega t}dt,
\end{eqnarray}
where $\eta_\omega=\mathrm{sgn}[\mathrm{Im}(\omega)]$.  %
Taking the absolute value of Eq.~\eqref{eq:G_XY_integral_rep} and using triangle inequality, we have
\begin{eqnarray}\label{eq:G_XY_LR}
	|\Omega_{XY}(\omega)|&\leq&\int^{\infty}_0|\langle G|[\hat{S}_X(t),\hat{V}_Y]|G\rangle| e^{-|\mathrm{Im}(\omega)| t}dt\nonumber\\
	&\leq&\int^{\infty}_0 C(d_{XY},t) e^{-|\mathrm{Im}(\omega)| t}dt\nonumber\\
	&\equiv&\bar{\Omega}(d_{XY},y),
\end{eqnarray}
where  $y=|\mathrm{Im}[\omega]|>0$. In the second line of Eq.~\eqref{eq:G_XY_LR} we assume a Lieb-Robinson-type bound $|\langle G|[\hat{S}_X(t),\hat{V}_Y]|G\rangle|\leq C(d_{XY},t)$, whose expression will be given in Secs.~\ref{sec:LPPL_small_alpha}-\ref{sec:LPPL_local} when we consider systems with different range of interaction. At large $t$, $C(d_{XY},t)$ equals the constant trivial bound $2\|S_X\| \|V_Y\|$, so $\bar{\Omega}(d_{XY},y)$ is finite for any $\omega$ with $\mathrm{Im}(\omega)\neq 0$, but diverges as $\bar{\Omega}(d_{XY},y)\sim 1/y$ when $y\to 0$, so gives no bound on the desired $|\Omega_{XY}(0)|$. %
Nevertheless, we can obtain a bound on $|\Omega_{XY}(0)|$ from the above by using a powerful technique from complex analysis. 
The analyticity of $\Omega_{XY}(\omega)$ allows us to improve the bound on $|\Omega_{XY}(\omega)|$ over the initial bound in Eq.~\eqref{eq:G_XY_LR}, by applying the following lemma~(Thm.~2.12 in Ref.~\cite{rosenblum1994topics}):
\begin{lemma}\label{thm:subharmonic}
	If $g(z)$ is complex analytic in a domain~(a simply connected open region) $S$, then $u(z)=\ln |g(z)|$ is a subharmonic function in $S$, i.e. for any $z_0\in S$ and $\rho>0$, if the circular region defined by $|z-z_0|\leq \rho$ is contained in $S$, then
	\begin{equation}\label{eq:subharmonic}
		u(z_0)\leq \frac{1}{2\pi}\int_0^{2\pi}u(z_0+\rho e^{i\theta}) d\theta.
	\end{equation}
\end{lemma}
Since $\Omega_{XY}(\omega)$ is complex analytic in the open disk region defined by $|\omega|< \Delta$,  Lemma~\ref{thm:subharmonic} implies
\begin{eqnarray}\label{eq:holographicbound}
	\ln |\Omega_{XY}(0)|&\leq& \frac{1}{2\pi}\int_0^{2\pi}\ln |\Omega_{XY}[\rho e^{i\theta}]| d\theta\nonumber\\
	&\leq & \frac{1}{2\pi}\int_0^{2\pi}\ln \bar{\Omega}[d_{XY},|\rho\sin\theta|] d\theta,
\end{eqnarray}
for any $\rho\in (0,\Delta)$. We will see that the integration over $\theta$ in the last line is convergent despite $\bar{\Omega}(d_{XY},y)$ diverging when $y\to 0$~\footnote{Notice that when $\theta=0$ or $\pi$, the function $\bar{\Omega}(d_{XY},|\rho\sin\theta|)$ in Eq.~\eqref{eq:holographicbound} is undefined. In a more rigorous treatment of Eq.~\eqref{eq:holographicbound}, we should first bound $\ln |\Omega_{XY}(0)|$ by $(\int_\epsilon^{\pi-\epsilon}+\int_{\pi+\epsilon}^{2\pi-\epsilon})\ln \bar{\Omega}[d_{XY},|\rho\sin\theta|] d\theta/2\pi +2\epsilon M_\rho/\pi$, where $M_\rho=\max_{|z|\leq \rho}\ln |\Omega_{XY}(z)|$ and the inequality holds for any $\epsilon>0$. So long as the improper integral in Eq.~\eqref{eq:holographicbound} converges, we can take the limit $\epsilon\to 0$ and obtain an upper bound for $\ln |\Omega_{XY}(0)|$ independent of $\epsilon$ and $M_\rho$, which is equal to the last line of Eq.~\eqref{eq:holographicbound}.}. 

The rest of our task is to insert the LRB of specific systems into Eq.~\eqref{eq:G_XY_LR} to get $\bar{\Omega}(d_{XY},y)$, and then compute the second line of Eq.~\eqref{eq:holographicbound} to obtain an upper bound for $|\Omega_{XY}(0)|$, which we  do in Secs.~\ref{sec:LPPL_large_alpha}-\ref{sec:LPPL_local}. In App.~\ref{app:conformal_mapping} we introduce a technique to further improve the bound in Eq.~\eqref{eq:holographicbound} using a conformal mapping.

\subsection{Power-law interactions with $\alpha>2D$}\label{sec:LPPL_large_alpha}
We start with the simplest case: $\alpha>2D$ and all interactions are two body, i.e. all the $h_X$ in Eq.~\eqref{eq:Hamiltonian} are of the form $h_X=h_{ij}V_iW_j$ where $V_i, W_j$ are local operators with unit norm and finite support separated by a distance $d_{ij}$ and $h_{ij}$ are real parameters satisfying $h_{ij}\leq \mathbf{C}d_{ij}^{-\alpha}$~\footnote{Note that this definition of two-body interactions is more general than the more common definition--here we actually allow interaction between any two clusters of sites, where the radius of each cluster is no larger than a fixed number. It is easy to see that the results in Ref.~\cite{Foss2015nearly} are still valid for two-cluster interactions, since one can always regroup lattice sites to make interactions two-body, as long as the cluster sizes have an upper bound.}. Similar to the $\mathbf{P}(x)$ notation, throughout this paper we use $\mathbf{C}$ to denote a positive constant independent of $r$ and $t$, and $\mathbf{C}$ in different equations or in different parts of the same equation need not be the same. %
In this case we use the Hastings-Koma bound~\cite{hastings2006} for short times, the algebraic light cone~\footnote{Although for the case $\alpha>2D+1$, a stronger linear light-cone has been obtained in Refs.~\cite{Lucas2019longrange,kuwahara2020strictly}, they do not give us qualitatively tighter LPPL bounds. Indeed, the LRB in Ref.~\cite{Lucas2019longrange} gives $C(r,t)\leq t/r$ in 1D, leading to an LPPL bound with $\alpha_1=1$, which is worse than our current results, and the LRB in Ref.~\cite{kuwahara2020strictly} can at most improve the subleading prefactor in Eq.~\eqref{eq:main_result}, since $\alpha_1=\alpha$ is already tight for generic systems.} Lieb-Robinson bound~\cite{Foss2015nearly} for intermediate times, and the trivial bound for long times
\begin{equation}\label{eq:powerlaw_LRB}
	C(r,t)\leq \begin{cases}
		\mathbf{C} e^{v't}/r^\alpha,& 0\leq t\leq t'\\
		\mathbf{C}e^{vt-\mathbf{C}\frac{r}{t^\gamma}}+\frac{\mathbf{C}t^{\alpha(1+\gamma)}}{r^\alpha},&t'< t\leq t_0,\\
		\mathbf{C}, &t>t_0,
	\end{cases} 
\end{equation}
where $t'=\alpha \ln \alpha /v$, $t_0=\mathbf{C}r^{1/(\gamma+1)}$, $v$ is a constant, and $\gamma=(1+D)/(\alpha-2D)$. 
Inserting these LRBs into Eq.~\eqref{eq:G_XY_LR} gives
\begin{eqnarray}\label{eq:Gry_large_alpha}
	\bar{\Omega}(r,y)&=&\int^{t'}_0 \mathbf{C}\frac{e^{(v'-y)t}}{r^\alpha}dt+\int^{t_0}_{t'} C(r,t) e^{-yt} dt\\
	&&+\int^{\infty}_{t_0} \mathbf{C} e^{-y t}dt\nonumber\\
	&\leq & \frac{\mathbf{C}t_0}{2}[e^{-\mathbf{C}r}+e^{(v-y)t_0-\mathbf{C}r/t_0^\gamma}] \nonumber\\
	&&{}+\frac{1}{r^\alpha}\left[\mathbf{C}+\mathbf{C}\frac{\Gamma[\alpha(\gamma+1)+1]}{y^{\alpha(\gamma+1)+1}}\right]+\mathbf{C}\frac{e^{-yt_0}}{y},\nonumber
\end{eqnarray}
where %
for the second term in the RHS of the first line we use Jensen's inequality since the integrand is convex when $t'\leq t\leq t_0$~(this convexity relies on a relation between the different constants here, which can always be satisfied, see App.~\ref{app:LPPL_largealpha}). The third line in Eq.~\eqref{eq:Gry_large_alpha} decays subexponentially in $r$, while the last line decays algebraically, so the term proportional to $r^{-\alpha}$ dominates the long-distance behavior of $\bar{\Omega}(r,y)$. Inserting $\bar{\Omega}(r,y)$ into Eq.~\eqref{eq:holographicbound}, one obtains Eq.~\eqref{eq:main_result}
where the subleading factor $\mathbf{P}(\ln r)$ is a constant in this case.  %
See App.~\ref{app:LPPL_largealpha} for details. 

\subsection{Power-law interactions with $\alpha>D$}\label{sec:LPPL_small_alpha}
The bound in the previous section does not apply to the case $D< \alpha<2D$, and is limited to two-body~(two-cluster) interactions. In this section we  consider the more general case where the $h_X$ in Eq.~\eqref{eq:Hamiltonian} satisfies~\cite{hastings2006} 
\begin{equation}\label{eq:power-law-condition}
	\sum_{X:X\supset \{i,j\}} \| h_X\|\leq \frac{h_0}{d_{ij}^\alpha},
\end{equation}
for all $i$ and $j$, with $\alpha>D$. In this case, the Hastings-Koma bound~\cite{hastings2006} is the tightest general LRB:
\begin{equation}\label{eq:Hastings-Koma}
\|[\hat{S}_X(t),\hat{V}_Y]\|\leq \min\{\mathbf{C}\frac{e^{vt}}{d_{XY}^\alpha},\mathbf{C}\}, 
\end{equation}
where $v$ is a positive constant. Using $|\langle G|[\hat{S}_X(t),\hat{V}_Y]|G\rangle|\leq \|[\hat{S}_X(t),\hat{V}_Y]\|$, and substituting $C(d_{XY},t)$ in Eq.~\eqref{eq:G_XY_LR} by the RHS of Eq.~\eqref{eq:Hastings-Koma} gives
\begin{eqnarray}\label{eq:Gry_smallalpha}
	\bar{\Omega}(r,y)&=&\int^{\infty}_0 C(r,t) e^{-y t}dt\nonumber\\
	&=&\int^{t_0}_0 \mathbf{C}\frac{e^{(v-y)t}}{r^\alpha} dt+\int^{\infty}_{t_0} \mathbf{C} e^{-y t}dt\nonumber\\
	&\leq & \mathbf{C} t_0\frac{1+e^{(v-y)t_0}}{2 r^\alpha}+ \mathbf{C} \frac{e^{-y t_0}}{y}\nonumber\\
	&=&\frac{\mathbf{C}t_0}{2r^\alpha}+\mathbf{C}\left(\frac{t_0}{2}+\frac{1}{y}\right)e^{-y t_0}\nonumber\\
	&\leq&\begin{cases}
	 \mathbf{C} t_0 e^{-y t_0}/y, &y\leq v,\\
		\mathbf{C} t_0 r^{-\alpha}, &y>v.
	\end{cases}
\end{eqnarray}
where we define $t_0=(\ln \mathbf{C}+\alpha\ln r)/v$, and in the third line we use Jensen's inequality~(due to the convexity of the integrand) to simplify the integral, rather than evaluating it exactly, in order to facilitate later computations. %

Now we insert Eq.~\eqref{eq:Gry_smallalpha} into Eq.~\eqref{eq:holographicbound}, to upper bound $|\Omega_{XY}(0)|$. %
Eq.~\eqref{eq:holographicbound} becomes
\begin{eqnarray}\label{eq:holographicbound_smallalpha_nonconformal}
	\ln |\Omega_{XY}(0)|&\leq& \ln Ct_0-\frac{2}{\pi}\int_{\theta_0}^{\pi/2}\alpha\ln r  d\theta\\
	&&-\frac{2}{\pi}\int_0^{\theta_0}(t_0\Delta\sin\theta+\ln \Delta\sin\theta)  d\theta,\nonumber
\end{eqnarray}
where $\theta_0=\arcsin(v/\Delta)$ if $v<\Delta$ and $\theta_0=\pi/2$ if $v\geq\Delta$. Finishing this integral, we obtain Eq.~\eqref{eq:main_result} with 
\begin{equation}\label{eq:alpha1_nonconformal}
	\alpha_1=\frac{2\alpha\Delta}{\pi v}(1-\cos\theta_0)+\alpha(1-\frac{2\theta_0}{\pi}).
\end{equation}
In App.~\ref{app:LPPL_smallalpha} we will improve this result using the technique of conformal mapping, and obtain the result in Tab.~\ref{tab:comp}. We will use the improved result~[Eq.~\eqref{eq:alpha1_smallalpha}] for the rest of this paper. 

\subsection{Short-range interacting systems}\label{sec:LPPL_local}
The method in Sec.~\ref{sec:improved_method} also significantly improves the LPPL bounds for systems with short range interactions, either exponentially decaying or strictly finite ranged. 
Specifically, we consider systems whose Hamiltonians Eq.~\eqref{eq:Hamiltonian} satisfy~\cite{hastings2006} 
\begin{equation}
	\sum_{X:X\supset \{i,j\}} \| h_X\|\leq h_0e^{-\mu d_{ij}},
\end{equation}
for all $i$ and $j$, where $\mu$ is some positive constant. The Lieb-Robinson bound is~\cite{hastings2006}
\begin{equation}\label{eq:LRB_exp_decay}
	C(r,t)\leq \mathbf{C} e^{-\mu (r-vt)}.
\end{equation}
Notice that the RHS of Eq.~\eqref{eq:LRB_exp_decay} can be obtained from the RHS of Eq.~\eqref{eq:Hastings-Koma} with the substitutions  $r\to e^r,  \alpha\to\mu, v\to \mu v$. We can therefore directly make this substitution in the results of Sec.~\ref{sec:LPPL_small_alpha} and App.~\ref{app:LPPL_smallalpha}, and obtain the bound
\begin{equation}\label{eq:GXY0_expdecay}
	|\Omega_{XY}(0)|\leq \mathbf{P}(r)e^{-\mu_1 r}
\end{equation}
with $\mu_1$ given in Tab.~\ref{tab:comp}. We see that for $\Delta\ll v$ our bound gives $\mu_1\approx \Delta/v$, which improves the previous best bound $\mu_1=\mu/(1+2\mu v/\Delta)\approx \Delta/(2v)$ by approximately a factor of 2.
Furthermore, %
if one wants a tighter bound for a specific model, one can use the LRB in Eq.~(32) of Ref.~\cite{Wang2020Tightening}: $C(r,t)\leq \mathbf{C} e^{\omega_m(i\kappa)t-\kappa r}, \forall \kappa>0$, where $\omega_m(i\kappa)$ is some (efficiently computable) function of $\kappa$~(Ref.~\cite{Wang2020Tightening} mainly deals with systems with finite range interactions, but the method can be directly generalized to systems with exponentially decaying interactions). %
This leads to a bound of the same form as Eq.~\eqref{eq:GXY0_expdecay} in which $\mu_1$ is a function of $\kappa$. One can then maximize $\mu_1(\kappa)$ over $\kappa>0$. This method gives further quantitative improvement for a specific model, especially at large $\Delta/v$.

\section{Generalization to gapped degenerate ground states}\label{sec:degenerate}
In this section we generalize our bounds to gapped systems with degenerate ground states. 
We begin with a straightforward extension. Notice that if the system has a subspace $\mathcal{H}_1\subseteq \mathcal{H}$ such that both the Hamiltonian $H$ and the perturbation $V_Y$ leave $\mathcal{H}_1$ invariant~(this is not required for $S_X$), and the ground state $|G_1\rangle$ of $\mathcal{H}_1$ is non-degenerate and gapped~(within $\mathcal{H}_1$), then all our proofs in the previous section applies to this subspace $\mathcal{H}_1$,  provided that $\bar{P}_G$ in Eq.~\eqref{eq:PTevolution} is understood as the projector to all the excited states within $\mathcal{H}_1$. In particular, if the system has a set of conserved quantum numbers that commute with both $H$ and $V_Y$ and  distinguish all the gapped degenerate ground states, then our bounds apply to all the ground states.  %

Nevertheless, this simple extension does not apply if the perturbation $V_Y$ breaks the conserved quantities. It also fails if the degeneracy is  not due to any symmetry at all, which includes the important class of topological degeneracy, where the~(approximately) degenerate ground states cannot be distinguished by local conserved quantum numbers. In the following we present a more  general treatment for degenerate ground states~(motivated by the method in Ref.~\cite{Hastings2007Goldstone}), which shows that all our results in Tab.~\ref{tab:comp} still hold provided that $\langle S_X\rangle$ is averaged over all the (nearly) degenerate ground states with equal weights. This can be thought of as the temperature $T\to 0$ limit of the statistical mechanical average, as long as this limit is taken after the thermodynamic limit $L\to \infty$, in which the splitting of ground state degeneracy vanishes.

Let us denote the degenerate ground states of $H(\lambda)=H+\lambda V_Y$ as $|G^a(\lambda)\rangle$,  with energy $E^a_0(\lambda)$, for $a=1,2,\ldots,d$, respectively. Notice that we do not require the degeneracy to be exact~(which is important for treating topological degeneracy), but only  that at each $\lambda$, all the ground state energies $E^a_0(\lambda)$ are separated from the rest of the spectrum~(the excited states) by at least an amount $\Delta(\lambda)>0$, and $\Delta(\lambda)$ is uniformly bounded from below, i.e. $\Delta\equiv\inf_{\lambda\in[0,1]}\Delta(\lambda)>0$. 
[Similar to the non-degenerate case, as long as $\Delta(0)>0$, the uniform gap condition is always satisfied for sufficiently small $\|V_Y\|$, as guaranteed by Weyl's inequality.]

The method follows Sec.~\ref{sec:improved_method}, but now using degenerate perturbation theory.
If some of the ground states are exactly degenerate at some $\lambda$, then we have some freedom to choose a basis for the exactly degenerate subspace, and it can be shown that~\cite{Hastings2007Goldstone} %
it is always possible to choose a suitable basis for this subspace such that $V_Y$ is  diagonal within this subspace and $\langle G^a(\lambda)|\partial_\lambda|G^b(\lambda)\rangle=0$ whenever $E_0^a(\lambda)=E_0^b(\lambda)$. Then degenerate perturbation theory generalizes Eq.~\eqref{eq:PTevolution} to 
\begin{equation}\label{eq:PTevolution_degen}
	\partial_\lambda|G^a(\lambda)\rangle=\frac{\bar{P}^a(\lambda)}{\hat{H}(\lambda)-E^a_0(\lambda)}V_Y|G^a(\lambda)\rangle,
\end{equation}
where 
\begin{eqnarray}\label{eq:projector_decomp}
\bar{P}^a(\lambda)&=&\mathds{1}-\sum_{b: E_0^b(\lambda)=E_0^a(\lambda)}|G^b(\lambda)\rangle\langle G^b(\lambda)|\\
&=&\bar{P}_G(\lambda)+\sum_{b: E_0^b(\lambda)\neq E_0^a(\lambda)}|G^b(\lambda)\rangle\langle G^b(\lambda)|,\nonumber
\end{eqnarray}
where $\bar{P}_G(\lambda)\equiv \hat{\mathds{1}}-\sum^d_{b=1}|G^b(\lambda)\rangle\langle G^b(\lambda)|$ is the projection operator to the space of all excited states.
Inserting the second line of Eq.~\eqref{eq:projector_decomp} into Eq.~\eqref{eq:PTevolution_degen}, we get 
\begin{equation}\label{eq:PTevolution_degen2}
	\partial_\lambda|G^a(\lambda)\rangle=\frac{\bar{P}_G(\lambda)}{\hat{H}(\lambda)-E^a_0(\lambda)}V_Y|G^a(\lambda)\rangle+\sum^d_{b=1}Q^{ab}|G^b(\lambda)\rangle,
\end{equation}
where 
\begin{equation}
	Q^{ab}=\begin{cases}
	\frac{\langle G^b(\lambda)|V_Y|G^a(\lambda)\rangle}{E_0^b(\lambda)-E_0^a(\lambda)},&\text{ if } E_0^b(\lambda)\neq E_0^a(\lambda),\\
	0, &\text{ if } E_0^b(\lambda)= E_0^a(\lambda),
	\end{cases}
\end{equation}
is an anti-Hermitian matrix $(Q^{ab})^*=-Q^{ba}$. 
We now consider the expectation value $\langle S_X\rangle_\lambda$ of a local observable $S_X$ averaged over all degenerate ground states $\{|G^b(\lambda)\rangle\}_{b=1}^d$, i.e. we define $\langle O\rangle_\lambda\equiv \frac{1}{d}\sum_{b=1}^d \langle G^b(\lambda)| O|G^b(\lambda)\rangle$ for any operator $O$. Then Eq.~\eqref{eq:deltaSX} becomes
\begin{equation}\label{eq:deltaSX_degen}
	\partial_\lambda\langle \hat{S}_X \rangle_{\lambda}=\left\langle\hat{S}_X\frac{\bar{P}_G(\lambda)}{\hat{H}(\lambda)-E_0^a(\lambda)}\hat{V}_Y\right\rangle_\lambda+\mathrm{c.c.},
\end{equation}
where, importantly, the contribution of the second term in Eq.~\eqref{eq:PTevolution_degen2} cancel due to anti-Hermiticity of $Q^{ab}$. The rest of  Sec.~\ref{sec:improved_method} generalizes in a straightforward way, with the only difference being that the ground state expectation value $\langle G(\lambda)|\ldots|G(\lambda)\rangle$ is replaced by the average $\langle\ldots\rangle_\lambda$. Lieb-Robinson bounds can still be used as we have $\langle[\hat{S}_X(t),\hat{V}_Y]\rangle_\lambda\leq \|[\hat{S}_X(t),\hat{V}_Y]\|\leq C(r,t)$. All resulting bounds remain the same as those listed in Tab.~\ref{tab:comp}.

\section{Implications for finite size numerical simulations}\label{sec:FSE}
\begin{figure}
	\center{\includegraphics[width=\linewidth]{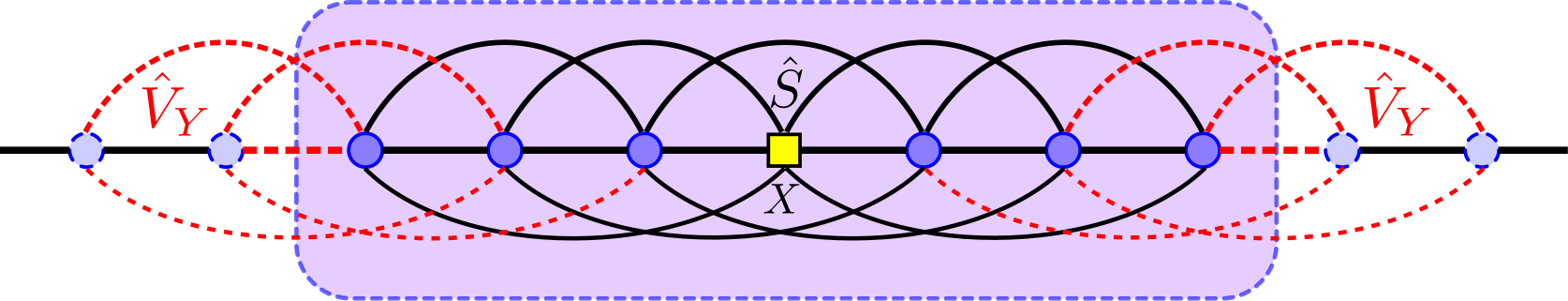}}
	\caption{\label{fig:FSE} Upper bounding finite-size error with the LPPL, illustrated for a 1D chain. The LPPL principle immediately gives an upper bound on finite size error of local observables in numerical simulation of gapped ground states, by recognizing $\hat{V}_Y$ as the interactions between the sites of the finite system and sites lying outside. 
	}
\end{figure}

In this section we present a straightforward application of our results, bounding the finite size errors~(FSEs) of local observables in gapped ground states of power-law systems, generalizing the bounds for locally-interacting systems proved in Ref.~\cite{Wang2021Bounding}. The basic configuration for the 1D case is illustrated in Fig.~\ref{fig:FSE}. The FSE for a local observable $\hat{S}_X$ measured in a $L$-site calculation is defined as $\delta\langle\hat{S}_X\rangle_L\equiv |\langle\hat{S}_X\rangle_L-\langle\hat{S}_X\rangle_\infty|$, which can be considered as the effect of the boundary interaction $\hat{V}_Y$ on $\hat{S}_X$, since removing $\hat{V}_Y$ from the thermodynamic Hamiltonian $\hat{H}$ decouples the finite system and the outside, leading to $\langle \hat{S}_X \rangle_{L}= \langle \hat{S}_X \rangle_{\hat{H}-\hat{V}_Y}$. We assume that the spectral gap $\Delta_L(\lambda)$ of the interpolated Hamiltonian  $\hat{H}-\lambda \hat{V}_Y$ is uniformly bounded from below $\min_{\lambda\in[0,1]}\Delta_L(\lambda) =\Delta>0$~\footnote{Actually, we only need to assume that there exists a gapped path $\hat{H}(\lambda)$ that connects $\hat{H}-\hat{V}_Y$ and $\hat{H}$. We do not require $\hat{H}(\lambda)$ to be a linear interpolation, but $\hat{H}(\lambda)$ should only differ from  $\hat{H}$ near the boundary.}.  
Under this assumption, we can apply our main result Eq.~\eqref{eq:main_result} to upper bound $\delta\langle\hat{S}_X\rangle_L$. A complication here is that $\hat{V}_{Y}$ contains infinitely many terms, including those that are very close to $\hat{S}_X$, so $r=d_{XY}$ is zero. To solve this issue, we can write 
\begin{equation}\label{eq:V_Y}
\hat{V}_{Y}=\sum_{i\in L,j\notin L}\hat{V}_{ij},
\end{equation}
where the summation is over all the interaction terms $\hat{V}_{ij}$ with $i$ in the $L$-site system and $j$ outside. Inserting Eq.~\eqref{eq:V_Y} into Eq.~\eqref{eq:deltaSX}, and using Eqs.~(\ref{eq:G_XY_spectral_decomp}-\ref{eq:holographicbound}) to upper bound the contribution of each individual $\hat{V}_{ij}$ term independently, we get%
\begin{equation}\label{eq:FSE_sum}
	|\delta\langle \hat{S}_X \rangle_{L}| \leq \sum_{i\in L,j\notin L} \|\hat{V}_{ij}\|\mathbf{P}(\ln r_{iX})/r_{iX}^{\alpha_1}.
\end{equation}
In the following we treat the 1D case for simplicity, and present the derivation in arbitrary dimension in App.~\ref{app:FSE_higherD}. Let $R=L/2$ and $\delta(r)=\mathbf{P}(\ln r)/r^{\alpha_1}$, we have 
\begin{eqnarray}\label{eq:FSE_sum_1D}
	|\delta\langle \hat{S}_X \rangle_{L}| &\leq& \sum_{-R \leq i\leq R,|j|> R}\delta(|i|+1)/(j-i)^\alpha\nonumber\\
	&\leq& \sum_{-R \leq i\leq R}\mathbf{C}\delta(|i|+1)/(R+1-i)^{\alpha-1}\nonumber\\
	&\leq &\sum^{R+1}_{i=1}\mathbf{P}(\ln i)i^{-\alpha_1}(R+2-i)^{1-\alpha}.
\end{eqnarray}
The following lemma gives a bound for the convolutional sum~(see App.~\ref{app:proof_power_law_convolution} for proof):
\begin{lemma}\label{lemma:power_law_convolution}
	Let $\eta,\zeta$ be real constants satisfying $0<\eta\leq \zeta$. Then 
	\begin{equation}\label{eq:lemma_convolution}
		\sum^{R-1}_{r=1}\frac{\mathbf{P}(\ln r)}{r^\zeta (R-r)^\eta }\asymp\mathbf{P}(\ln R)\times\begin{cases}%
			R^{-\eta},&\text{ if }\zeta\geq1,\\
			R^{1-\eta-\zeta},&\text{ if } \zeta<1,%
		\end{cases}
	\end{equation}
where the notation $f(R)\asymp g(R)$ means that there exist positive constants $c_1,c_2$ independent of $R$ such that  $c_1 g(R)\leq f(R)\leq c_2 g(R)$ for all $R\in \mathbb{Z}_{\geq 1}$.
\end{lemma}
Applying Lemma~\ref{lemma:power_law_convolution} to Eq.~\eqref{eq:FSE_sum_1D}, we obtain Eq.~\eqref{eq:FSEB} with 
$$\alpha_3=\begin{cases}
	\alpha_1+\alpha-2 & \text{\!\!\!if } \alpha_1\leq 1\\
	\alpha-1   &\text{\!\!\!if } \alpha_1>1
\end{cases}$$
for $1<\alpha\leq 2$, and $\alpha_3=\alpha-1$ for $\alpha>2$,  which is the result in Tab.~\ref{tab:comp} for $D=1$. 
\section{Improved bounds on ground state correlation decay}\label{sec:correlation}
In this section we show that the method in Sec.~\ref{sec:improved_method} also significantly improves bounds on  correlation decay of gapped~(possibly degenerate) ground states, compared to previous results~\cite{hastings2006,nachtergaele2006}.
We first obtain an integral formula that relates $\Omega_{XY}(\omega)$ in Eq.~\eqref{eq:G_XY_spectral_decomp} and the connected correlation function $\langle S_X V_Y\rangle_c\equiv \langle S_X V_Y\rangle-\langle S_X\rangle\langle V_Y\rangle $ in the gapped ground state $|G\rangle$. Integrating  Eq.~\eqref{eq:G_XY_spectral_decomp} along the imaginary axis, we have 
\begin{eqnarray}\label{eq:G_XY_omega_integral}
	\int^{+\infty i}_{-\infty i}\Omega_{XY}(\omega)d\omega
	&=&\int^{+\infty i}_{-\infty i} d\omega\langle G|\hat{S}_X\frac{i\bar{P}_G}{\omega-\hat{\Delta}}\hat{V}_Y|G\rangle\nonumber\\
	&&-\int^{+\infty i}_{-\infty i} d\omega\langle G|\hat{V}_Y\frac{i\bar{P}_G}{\omega+\hat{\Delta}}\hat{S}_X|G\rangle\nonumber\\
	&=&\pi  \langle G|\hat{S}_X\bar{P}_G\hat{V}_Y|G\rangle+\mathrm{c.c.}\nonumber\\
	&=&2\pi \langle S_X V_Y\rangle_c,
\end{eqnarray}
where we used the following equality
\begin{equation}\label{eq:omega_integral}
	\int^{+\infty i}_{-\infty i}\frac{1}{\omega-\mu}d\omega= -\pi i ~\mathrm{sgn}(\mu).
\end{equation}
With Eq.~\eqref{eq:G_XY_omega_integral}, we can obtain an upper bound on $|\langle S_X V_Y\rangle_c|$ by integrating $|\Omega_{XY}(\omega)|$ along the imaginary axis. Furthermore, it can be proved that~(see App.~\ref{app:MaxOmega}) $|\Omega_{XY}(\omega)|$ on the imaginary axis can always be upper bounded by the upper bound of $|\Omega_{XY}(0)|$ obtained by Eq.~\eqref{eq:holographicbound}~(we denote this upper bound by $|\bar{\Omega}_{XY}(0)|$). Therefore, we can use the upper bound $|\Omega_{XY}(i y)|\leq \min[|\bar{\Omega}_{XY}(0)|,\bar{\Omega}(d_{XY},y)]$. Notice that the integration on this bound on $iy$ is guaranteed to converge provided one uses the best LRB, since $C(d_{XY},t)\propto t^\nu$ at small $t$ with $\nu\geq 1$, and so $\bar{\Omega}(d_{XY},y)$ in  Eq.~\eqref{eq:G_XY_LR} decays at least as $y^{-\nu-1}$ at large $y$. This upper bound yields
\begin{eqnarray}\label{eq:SXVYbound}
	2\pi |\langle S_X V_Y\rangle_c|&\leq& 	\int^{+\infty }_{-\infty }|\Omega_{XY}(iy)|dy\\
	&\leq& 2y_0 |\bar{\Omega}_{XY}(0)| +2\int^\infty_{y_0}\bar{\Omega}(d_{XY},y)dy,\nonumber
\end{eqnarray}
for any $y_0>0$~[for the optimal result, $y_0$ should satisfy $\bar{\Omega}(d_{XY},y_0)=|\bar{\Omega}_{XY}(0)|$]. 

For example, for $D<\alpha<2D$, we have
 \begin{equation}\label{eq:Omega(r,y)_smallalpha}
	\bar{\Omega}(r,y)\leq \frac{\mathbf{C}}{r^{\alpha}}\left[\frac{e^{(v-y)t_0}-1}{v-y}+\frac{e^{-yt_0}-1}{y}\right]+ \mathbf{C}\frac{e^{-y t_0}}{y}.
\end{equation}
Inserting Eq.~\eqref{eq:Omega(r,y)_smallalpha} into Eq.~\eqref{eq:SXVYbound} and taking $y_0=v$, we see that the integral of the term in the square bracket converges to a constant independent of $r$, and therefore the second term in the last line of Eq.~\eqref{eq:SXVYbound} is bounded by $\mathbf{C}/d_{XY}^\alpha$.
For $|\bar{\Omega}_{XY}(0)|$ we use the result of %
 App.~\ref{app:LPPL_smallalpha}~[Eqs.~\eqref{eq:GXYb_smallalpha} and \eqref{eq:alpha1_smallalpha}]. In the end we obtain 
\begin{equation}\label{eq:corr_decay_powerlaw}
|\langle S_X V_Y\rangle_c| \leq \mathbf{P}(\ln d_{XY})/d_{XY}^{\alpha_1},
\end{equation}
where $\mathbf{P}(x)$ is a quadratic polynomial in $x$. 
Other cases in Tab.~\ref{tab:comp} can be treated in an identical manner, by inserting the results of Sec.~\ref{sec:main} into Eq.~\eqref{eq:SXVYbound}. In all cases, one obtains Eq.~\eqref{eq:correlation_decay_bound} with $\alpha_2=\alpha_1$ for the power-law cases or $\mu_2=\mu_1$ for short-range interacting cases. 

\section{Conclusion}\label{sec:conclusions}
We have proved a locality principle for  gapped ground states in systems with power-law~($1/r^\alpha$) decaying interactions: 
when $\alpha>D$, the response of a local observable $S_X$ to a spatially separated local perturbation $V_Y$ decays as a power-law~($1/r^{\alpha_1}$) in distance, 
provided that $V_Y$ does not close the spectral gap. When $\alpha>2D$, the bound on the exponent $\alpha_1$ that we obtain, $\alpha_1=\alpha$, is tight. We proved this using a method that avoids the use of QAC and incorporates techniques of complex analysis. 
Our method also  improves bounds on ground state correlation decay, even in short-range interacting systems. 

Our results have profound significance in studying the ground state properties of power-law interacting systems.  
At a fundamental level, the LPPL bounds generalize the  notion of locality to gapped ground states of power-law  systems, implying that the local properties of such ground states are stable against distant local perturbations. 
At a more practical level, we showed how our results immediately lead to an upper bound on finite size error in numerical simulations of gapped ground states, which revealed that FSEs generally decay as a power-law~($1/L^{\alpha_3}$) in system size~(provided that $\alpha$ or the spectral gap $\Delta$ is not too small). A corollary of this is the existence of  thermodynamic limit for local observables in ground states of power-law systems, under the spectral gap assumption stated in Sec.~\ref{sec:FSE}. 

We now discuss some open questions and future directions. One open question concerns whether the power law exponents $\alpha_1$ and $\alpha_2$ given in Tab.~\ref{tab:comp} are tight when $D<\alpha<2D$: we see that in this case both of them are strictly smaller than $\alpha$, yet for all gapped power-law systems we know, no correlations decay slower than $1/r^\alpha$, which strongly suggests that our bounds can further be improved in this case. 
An interesting future direction is to generalize our results to systems of interacting bosons, such the Bose-Hubbard model, where our current bounds do not apply due to the interaction $h_X$ in Eq.~\eqref{eq:Hamiltonian} having infinite norm, thereby violating Eq.~\eqref{eq:power-law-condition} and the corresponding LRBs. However,   our method in Sec.~\ref{sec:improved_method} still works if we incorporate Eq.~\eqref{eq:G_XY_LR} with recent LR-type bounds for interacting bosons~\cite{schuch2011information,Kuwahara2021InteractingBoson,Yin2022Finite,Faupin2022Maximal}. It will then be interesting to see how the exponents in Tab.~\ref{tab:comp} get modified.
Another future direction is to prove the stability of the spectral gap against extensive local perturbations in gapped frustration-free ground states of power-law Hamiltonians. For locally interacting systems, this has been proved under the local topological quantum order condition~\cite{bravyi2010topological,TPstability,michalakis2013stability}, where an essential tool in the proof is Hastings' QAC~[Eq.~\eqref{eq:QAC}]. It is interesting to investigate if our new method can improve these results and extend them to power-law systems. %

\acknowledgments
We thank Jens Eisert, Andrew Guo, Simon Lieu, and Alexey Gorshkov  for discussions. 
This work was supported in part with funds from the Welch Foundation~(C-1872), National Science Foundation~(PHY-1848304), and Office of Naval Research~(N00014-20-1-2695). KRAH thanks the Aspen Center for Physics, supported by the National Science Foundation grant PHY-1066293, and the KITP, supported in part by the National Science Foundation under Grant PHY-1748958,  for their hospitality while part of this work was performed.

\appendix
\section{LPPL bound from QAC}\label{appen:LPPL_QAC}
In this appendix we briefly show how to obtain a LPPL bound by directly generalizing the previous method based on QAC. We will see that the QAC only leads to a LPPL bound for $\alpha>2D$, where it gives $\alpha_1=\alpha-D-1$,  much looser than our bound $\alpha_1=\alpha$~(which we know to be tight).

We use the same set-up as in Sec.~\ref{sec:main}. The QAC constructs a unitary evolution process relating the ground states of different $\lambda$
\begin{equation}
	i\partial_\lambda |G(\lambda)\rangle=D(\lambda)|G(\lambda)\rangle,
\end{equation}
where $D(\lambda)$ is a Hermitian operator that depends on $H(\lambda)$. Following the derivations in Eqs.~(5-9) in Ref.~\cite{Gong2017Entanglement},  $D(\lambda)$ can be expanded as
\begin{equation}
	D(\lambda)=\sum_{R=1}^\infty V_Y(\lambda,R),
\end{equation}
where $V_Y(\lambda,R)$ is an operator that acts only on sites within a distance $R$ from $Y$. For $\alpha>2D$, $\|V_Y(\lambda,R)\|\leq \mathbf{C}\|V_Y\|/R^{\alpha-D}$, while for $\alpha\leq 2D$, $\|V_Y(\lambda,R)\|$ decays slower than any power in $R$~\cite{Gong2017Entanglement}.
We now use the previous method~\cite{hastings2010locality,LPPL} to bound $|\delta\langle \hat{S}_X \rangle_{\hat{V}_Y}|$:
\begin{eqnarray}\label{def:delta_SX_QAC}
	|\delta\langle \hat{S}_X \rangle_{\hat{V}_Y}|&\equiv& |\langle \hat{S}_X \rangle_{\lambda=1}-\langle \hat{S}_X \rangle_{\lambda=0}|\nonumber\\
	&\leq& \int_0^1 |\partial_\lambda \langle \hat{S}_X \rangle_{\lambda}|d\lambda\nonumber\\
	&=& \int_0^1 | \langle [\hat{S}_X,D(\lambda)] \rangle_{\lambda}|d\lambda\nonumber\\
	&\leq&\int_0^1 \| [\hat{S}_X,D(\lambda)] \|d\lambda.
\end{eqnarray}
For $\alpha>2D$, the integrand is bounded as
\begin{eqnarray}
	\|[\hat{S}_X, D(\lambda)]\|&=&\sum_{R\geq d_{XY}}\|[\hat{S}_X, V_Y(\lambda,R)]\|\nonumber\\
	&\leq&\sum_{R\geq d_{XY}} \mathbf{C}\frac{1}{R^{\alpha-D}}\nonumber\\
	&=&\mathbf{C}\frac{1}{d_{XY}^{\alpha-D-1}}.
\end{eqnarray}
This leads to the bound $|\delta\langle \hat{S}_X \rangle_{\hat{V}_Y}|\leq \mathbf{C}/d_{XY}^{\alpha-D-1}$ for $\alpha>2D$, while for $\alpha\leq 2D$, the bound obtained this way decays slower than any power in $d_{XY}$, verifying our earlier claims.

\section{Some details for Sec.~\ref{sec:main}}\label{app:LPPLdetails}
In this appendix we provide some technical details for Sec.~\ref{sec:main}. 
\subsection{Details for Sec.~\ref{sec:LPPL_large_alpha}}\label{app:LPPL_largealpha}
We first briefly explain how Eq.~\eqref{eq:powerlaw_LRB} is obtained from Ref.~\cite{Foss2015nearly}. The main result of Ref.~\cite{Foss2015nearly} is stated in their Eq.~(18)
\begin{equation}\label{eq:powerlaw_LRB_app}
	C(r,t)\leq \mathbf{C}\exp\left(vt-\frac{r}{\chi}\right)+\mathbf{C}\frac{e^{v_\chi t}}{[r/R(t)]^\alpha},
\end{equation}
valid when $vt>\alpha \ln \alpha$ and $r>6R(t)$, where $R(t)=\chi v t$, $v_\chi=\mathbf{C} R(t)^D\lambda_\chi$, and  $\lambda_\chi=\sup_{i\in \Lambda}\sum_{j:d_{ij}\geq \chi}\|h_{ij}\|$. For $\chi>\mathbf{C}$, we have $\lambda_\chi\leq \mathbf{C} \chi^{D-\alpha}$. We now take $\chi=C_0 t^\gamma$ where $C_0>0$ is a constant, so $R(t)=C_0 v t^{\gamma+1}$, and $r>6R(t)$ is equivalent to $t<(r/6vC_0)^{1/(\gamma+1)}\equiv t_0$. For $t>(\alpha\ln\alpha) /v$, we have $\chi>\mathbf{C}$, leading to $v_\chi t\leq \mathbf{C}$. Inserting $R(t)$ and $\chi$ into Eq.~\eqref{eq:powerlaw_LRB_app}, we get Eq.~\eqref{eq:powerlaw_LRB}. By taking the second derivative~(with respect to $t$) of the first term in the RHS of Eq.~\eqref{eq:powerlaw_LRB_app}, we see that this term is indeed convex provided that  the constant $C_0$ is chosen to be large enough, verifying our claim below Eq.~\eqref{eq:Gry_large_alpha}.

We now insert Eq.~\eqref{eq:Gry_large_alpha} into Eq.~\eqref{eq:holographicbound} to prove Eq.~\eqref{eq:main_result}. We first simplify the last line of Eq.~\eqref{eq:Gry_large_alpha}: notice that for $y=|\mathrm{Im}[\omega]|=\Delta|\sin \theta|\leq \Delta$, we have $e^{(v-y)t_0- r/(C_0t_0^\gamma)}\leq\mathbf{C} e^{-yt_0}/y$, $r^{-\alpha} \leq \mathbf{C}r^{-\alpha} y^{-\alpha(\gamma+1)-1}$, and $t_0 e^{-\mathbf{C}r}\leq \mathbf{C}r^{-\alpha} y^{-\alpha(\gamma+1)-1}$~(for $r\geq 1$). Therefore
\begin{equation}\label{eq:Gry_large_alpha_app}
	\bar{\Omega}(r,y)\leq (\mathbf{C}t_0+\mathbf{C}y^{-1})e^{-yt_0}
	+\mathbf{C}r^{-\alpha} y^{-\alpha(\gamma+1)-1}.
\end{equation}
The second term in Eq.~\eqref{eq:Gry_large_alpha_app} dominates at small and large $y$, while the first term is only important in an intermediate region $(y_1, y_2)$, where  $y_{1,2}=x_{1,2} r^{-1/(\gamma+1)}$ and $x_1,x_2$ are the two solutions to the equation~(and are independent of $r$)
\begin{equation}\label{eq:separation_point}
	(x+\mathbf{C})e^{-\mathbf{C}x}=x^{-\alpha(\gamma+1)}.
\end{equation} 
In summary, 
\begin{equation}\label{eq:Gry_large_alpha_app2}
	\bar{\Omega}(r,y)\leq
	\begin{cases}
		 (\mathbf{C}t_0+\mathbf{C}y^{-1})e^{-yt_0}, &y_1\leq y\leq y_2\\
	\mathbf{C}r^{-\alpha} y^{-\alpha(\gamma+1)-1}, &0<y<y_1\text{ or }y>y_2.
	\end{cases}
\end{equation}
[In case Eq.~\eqref{eq:separation_point} has no solution, then $\bar{\Omega}(r,y)$ is always bounded by the second line of Eq.~\eqref{eq:Gry_large_alpha_app2}, and our following derivations still work with minor modifications.] 
Inserting Eq.~\eqref{eq:Gry_large_alpha_app2} into Eq.~\eqref{eq:holographicbound}, we have
\begin{eqnarray}\label{eq:holographicbound_app}
	\ln |\Omega_{XY}(0)|&\leq& \ln \mathbf{C}-\frac{2\alpha}{\pi}(\pi/2-\theta_2+\theta_1)\ln r\\
	&&+\frac{2}{\pi}\int_{\theta_1}^{\theta_2}\left[\ln\left(\mathbf{C}t_0+\frac{\mathbf{C}}{\sin\theta}\right)-t_0\Delta\rho\sin\theta\right]d\theta,\nonumber%
\end{eqnarray}
where $y_{1,2}\equiv \Delta\sin\theta_{1,2}=x_{1,2} r^{-1/(\gamma+1)}$. Using $\theta_{1,2}=O[r^{-1/(\gamma+1)}]$, we see that all but the $\ln \mathbf{C}-\alpha\ln r$ term are of order $r^{-1/(\gamma+1)}$, $r^{-1/(\gamma+1)}\ln r$, or $r^{-2/(\gamma+1)}$, all of which are upper bounded by a constant for $r\geq 1$.  %
This proves Eq.~\eqref{eq:main_result}
with the subleading factor $\mathbf{P}(\ln r)$ being a constant.

\subsection{Improving the bound on $|\Omega_{XY}(0)|$ with conformal mapping}\label{app:conformal_mapping}
We now introduce a technique to further improve the bound in Eq.~\eqref{eq:holographicbound}, which leads to an improvement of the bound on $\alpha_1$ in Sec.~\ref{sec:LPPL_small_alpha}. The basic idea is to apply a suitable conformal mapping to $\Omega_{XY}(\xi)$ before applying the bound Eq.~\eqref{eq:holographicbound}. To be specific, let $f(\xi)$ be a complex analytic function in the open unit disk $D$, such that $f(0)=0$ and $f(D)\cap K_\Delta=\emptyset$.  
Then $\Omega_{XY}[f(\xi)]$ is complex analytic for $\xi\in D$, so according to Lemma~\ref{thm:subharmonic}, $\ln |\Omega_{XY}[f(\xi)]|$ is subharmonic in $D$, and therefore for any $\rho\in (0,1)$,
\begin{eqnarray}\label{eq:holographicbound_conformal}
	\ln |\Omega_{XY}(0)|&\leq& \frac{1}{2\pi}\int_0^{2\pi}\ln |\Omega_{XY}[f(\rho e^{i\theta})]| d\theta\\
	&\leq & \frac{1}{2\pi}\int_0^{2\pi}\ln \bar{\Omega}[d_{XY},|\mathrm{Im}f(\rho e^{i\theta})|] d\theta\nonumber.
\end{eqnarray}
Notice that Eq.~\eqref{eq:holographicbound} in Sec.~\ref{sec:improved_method} corresponds to the special case $f(\xi)=\Delta\xi$. Since the inequality~\eqref{eq:holographicbound_conformal} holds for all such functions $f(\xi)$~(satisfying the conditions mentioned above), we can choose a $f(\xi)$ to optimize this bound. We will show in the next section how this additional conformal mapping improves the bound in Sec.~\ref{sec:LPPL_small_alpha}.

\subsection{Improving the bound in Sec.~\ref{sec:LPPL_small_alpha}}\label{app:LPPL_smallalpha}
\begin{figure}
	\center{\includegraphics[width=\linewidth]{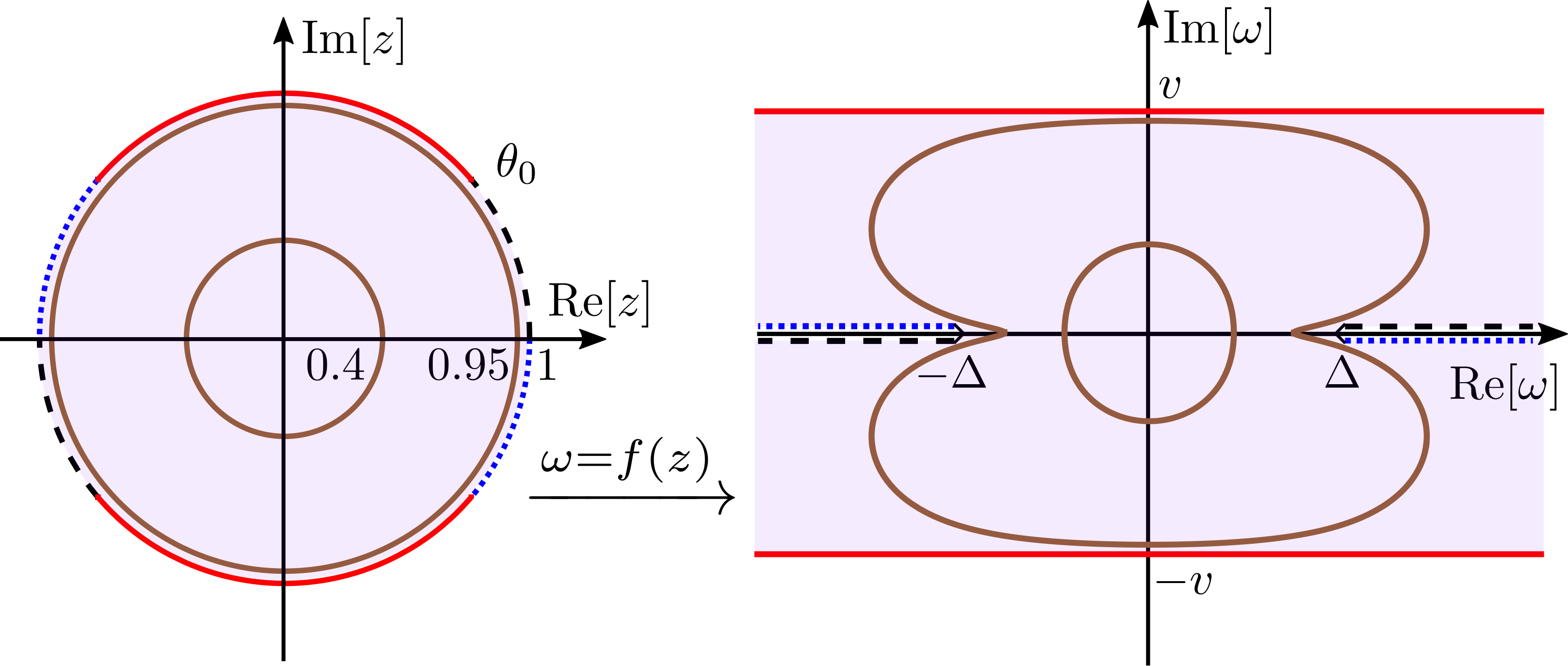}}
	\caption{\label{fig:omega_plane} For any finite system size $L$, $K_\Delta$~(dashed region on the real axis of the right panel) contains all possible pole positions of the RHS of Eq.~\eqref{eq:G_XY_spectral_decomp}, so $\Omega_{XY}(\omega)$ is complex analytic in the region $\mathbb{C}\backslash K_\Delta$. 
		The conformal mapping $\omega=f(z)$ defined in Eq.~\eqref{eq:g_v_conformal} maps the unit disk~(left) to the shaded region of the infinite strip with the pole regions excluded~(right).
	}
\end{figure}
We begin by inserting Eq.~\eqref{eq:Gry_smallalpha} into Eq.~\eqref{eq:holographicbound_conformal}, with the conformal mapping~\footnote{This conformal mapping is motivated by the answer to our question on math stackexchange https://math.stackexchange.com/questions/4443310/a-boundary-value-problem-of-a-harmonic-potential. We thank the user named ``messenger'' for providing the answer. } 
\begin{equation}\label{eq:g_v_conformal}
	f(z)\equiv\frac{2v}{\pi}\arctanh\left(\frac{2z}{ z^2+1}\tanh\frac{\Delta\pi}{2v}\right).
\end{equation}
The image of the unit disk under the mapping $\omega=f(z)$ is shown in Fig.~\ref{fig:omega_plane}. Notice that  $y(\theta)\equiv|\mathrm{Im}[f(\rho e^{i\theta})]|<v$ for $\rho\in (0,1),\theta\in [0,2\pi]$, so Eq.~\eqref{eq:holographicbound_conformal} becomes
\begin{eqnarray}\label{eq:holographicbound_smallalpha}
	\ln |\Omega_{XY}(0)|&\leq& \frac{2}{\pi}\int_0^{\pi/2}[\ln (\mathbf{C}t_0)-\ln y(\theta)- y(\theta)t_0] d\theta\nonumber\\
	&=&\ln( \mathbf{C}t_0)-\frac{2}{\pi}\int_0^{\pi/2}[\ln y(\theta)+y(\theta)t_0] d\theta.\nonumber\\
\end{eqnarray}
Before going into more technical calculations, we first give some heuristic arguments about the asymptotic behavior of the $|\Omega_{XY}(0)|$ at large $r$, and guess the exponent $\alpha_1$. We will see later that the asymptotic behavior of the last line of Eq.~\eqref{eq:holographicbound_smallalpha} at large $r$ is dominated by the third term, since the first two terms have much weaker dependence on $r$. The third term in the last line of Eq.~\eqref{eq:holographicbound_smallalpha} decreases as $\rho$ gets closer to $1$, and in the limit $\rho\to 1$, $y(\theta)$ becomes a step function: $y(\theta)=0$ for $\theta<\theta_0$ while $y(\theta)=v$ for $\theta>\theta_0$, where $\theta_0$ satisfies $\cos\theta_0=\tanh\frac{\Delta\pi}{2v}$ and is marked in Fig.~\ref{fig:omega_plane}. Therefore in the limit $\rho\to 1$ the third term in the last line of Eq.~\eqref{eq:holographicbound_smallalpha} is 
\begin{eqnarray}\label{eq:b=1limit}
	-t_0\frac{2}{\pi}\int_0^{\pi/2}y d\theta&=&-2t_0v (\pi/2-\theta_0)/\pi\nonumber\\
	&=&-\frac{2t_0v}{\pi} \arcsin\left(\tanh\frac{\Delta\pi}{2v}\right).
\end{eqnarray}
The subtlety here is that the first two terms in the last line of Eq.~\eqref{eq:holographicbound_smallalpha} diverges as $\rho\to 1$. In the following we will show that by choosing $\rho$ suitably close to $1$, we can obtain a bound
\begin{equation}\label{eq:GXYb_smallalpha}
	|\Omega_{XY}(0)|\leq P(\ln r)r^{-\alpha_1},
\end{equation}
with $\mathbf{P}(x)$ being a quadratic polynomial in $x$ and
\begin{equation}\label{eq:alpha1_smallalpha}
\alpha_1=\frac{2\alpha}{\pi}\arcsin\left(\tanh\frac{\Delta\pi}{2v}\right).
\end{equation}

We begin by upper bounding the first term in the integrand in Eq.~\eqref{eq:holographicbound_smallalpha}. Due to the symmetry $y(\theta)=y(\pi-\theta)=y(\pi+\theta)$, we only need to treat the integrand in the interval $\theta\in [0,\pi/2]$. To this end, we obtain a simple lower bound for $y(\theta)$ as follows:
\begin{eqnarray}
	y(\theta)&=&\frac{2v}{\pi}\mathrm{Im}\left[\arctanh\left(\frac{2z}{ z^2+1}\tanh\frac{\Delta\pi}{2v}\right)\right]\nonumber\\
	&\geq&\frac{2v}{\pi}\arctan\left[\mathrm{Im}\left(\frac{2z}{ z^2+1}\right)\tanh\frac{\Delta\pi}{2v}\right]\nonumber\\
	&=&\mathbf{C}\arctan\left[\mathbf{C}\frac{(1-\rho^2)\rho\sin\theta}{\rho^4+2\rho^2\cos2\theta+1}\right]\nonumber\\
	&\geq& \mathbf{C}(1-\rho)\sin\theta,
\end{eqnarray}
for $\rho\geq 0.9$ and $\theta\in [0,\pi/2]$, where in the second line we used $\mathrm{Im}[\arctanh
z]\geq \arctan\mathrm{Im}[z]$ for $z$ in the upper half plane~(which follows from the fact that $\mathrm{Im}[\arctanh (x+i \epsilon)]$ is monotonically increasing in $x$ for $\epsilon>0,x>0$), and the proof for the last line is elementary. %
Therefore the second term in the last line of Eq.~\eqref{eq:holographicbound_smallalpha} can be upper bounded by
\begin{eqnarray}\label{eq:integrand1}
	-\frac{2}{\pi}\int_0^{\pi/2}\ln y(\theta) d\theta\leq \ln \mathbf{C}-\ln (1-\rho).
\end{eqnarray}
This may be a crude  bound, but it captures the leading singularity of this term as $\rho\to 1$. We now study the second term in the integrand in Eq.~\eqref{eq:holographicbound_smallalpha} near $\rho\to 1$. We have
\begin{eqnarray}
	\partial_\rho y(\theta)&=&\mathrm{Im}[\partial_\rho f(z)]\nonumber\\
	&=&\mathrm{Im}\left[\frac{1}{i\rho }\partial_\theta f(z)\right]\nonumber\\
	&=&-\frac{1}{\rho}\mathrm{Re}[\partial_\theta f(z)],
\end{eqnarray}
and therefore
\begin{eqnarray}
	\partial_\rho \int_0^{\pi/2}y(\theta) d\theta&=& -\frac{1}{\rho}\int_0^{\pi/2}d\theta \partial_\theta \mathrm{Re}[f(z)] \nonumber\\
	&=& -\frac{1}{\rho}\mathrm{Re}[f(i\rho)-f(\rho)] \nonumber\\
	&=& \frac{1}{\rho}f(\rho),
\end{eqnarray}
the limit of which at $\rho\to 1$ is $\Delta$. Since this derivative exists for $\rho\in [\rho_0,1]$ for any $\rho_0>0$, along with Eq.~\eqref{eq:b=1limit}, we obtain
\begin{eqnarray}\label{eq:integrand2}
	\int_0^{\pi/2}y d\theta
	\geq C_0-\mathbf{C}(1-\rho),
\end{eqnarray}
where $C_0\equiv v\arcsin[\tanh(\Delta\pi/2v)]$.
Inserting Eqs.~(\ref{eq:integrand1},\ref{eq:integrand2}) into Eq.~\eqref{eq:holographicbound_smallalpha}, we get 
\begin{equation}\label{eq:holographicbound_smallalpha_final}
	\ln |\Omega_{XY}(0)|\leq -\alpha_1\ln r+\ln\frac{ \mathbf{C}t_0}{1-\rho}+\mathbf{C}t_0(1-\rho).%
\end{equation}
Minimizing the RHS of Eq.~\eqref{eq:holographicbound_smallalpha_final}~[the minimum is at $1-\rho=1/(\mathbf{C}t_0)$], we get Eq.~\eqref{eq:GXYb_smallalpha} where the polynomial prefactor can be taken as $P(\ln r)=\mathbf{C}(\ln r)^2$.

Comparing Eq.~\eqref{eq:alpha1_smallalpha} and Eq.~\eqref{eq:alpha1_nonconformal}, we see that the technique here improves $\alpha_1$ at all values of $\Delta/v$, especially when $\Delta/v$ is large, where $\alpha_1$ approaches $\alpha$ exponentially fast in Eq.~\eqref{eq:alpha1_smallalpha}, while $\alpha-\alpha_1\propto v/\Delta$ in Eq.~\eqref{eq:alpha1_nonconformal}.

\section{Finite-size error bounds}\label{app:FSEB}
In this appendix section we provide some missing details in Sec.~\ref{sec:FSE}, including a proof of Lemma~\ref{lemma:power_law_convolution} and a  derivation of the bounds in arbitrary spatial dimension. 
\subsection{Proof of Lemma~\ref{lemma:power_law_convolution}}\label{app:proof_power_law_convolution}
	For simplicity we assume $R=2R_1+1$ is an odd number~(the proof for even $R$ is similar).  We have %
	\begin{eqnarray}\label{eq:lemma_convolution_proof}
		\sum^{R-1}_{r=1}\frac{\mathbf{P}(\ln r)}{r^\zeta (R-r)^\eta }&=&\left(\sum^{R_1}_{r=1}+\sum^{R-1}_{r=R_1+1}\right)\frac{\mathbf{P}(\ln r)}{r^\zeta (R-r)^\eta }\nonumber\\
		&=&\sum^{R_1}_{r=1}\left\{\frac{\mathbf{P}(\ln r)}{r^\zeta (R-r)^\eta }+\frac{\mathbf{P}[\ln (R-r)]}{r^\eta (R-r)^\zeta }\right\}\nonumber\\
		&\asymp&\sum^{R_1}_{r=1}\left[\frac{\mathbf{P}(\ln r)}{r^\zeta R^\eta }+\frac{\mathbf{P}(\ln R)}{r^\eta R^\zeta }\right],\nonumber\\
		&\leq&\sum^{R_1}_{r=1}\left[\frac{\mathbf{P}(\ln R)}{r^\zeta R^\eta }+\frac{\mathbf{P}(\ln R)}{r^\eta R^\zeta }\right],
	\end{eqnarray}
	where in the second line we substituted $r$ by $R-r$ in the second sum, and in the third line we used $\mathbf{P}[\ln (R-r)](R-r)^{-\gamma}\asymp \mathbf{P}(\ln R)R^{-\gamma}$ for $1\leq r\leq R_1$ and $\gamma>0$ since $\mathbf{P}[\ln (R/2)]\leq\mathbf{P}[\ln (R-r)]\leq \mathbf{P}(\ln R) $ and $R^{-\gamma}\leq (R-r)^{-\gamma}\leq (R/2)^{-\gamma}$. %
	Now applying $\sum^{R_1}_{r=1}r^{-\gamma}\asymp \int^{R_1}_{r=1}r^{-\gamma} dr$ to the last line  of Eq.~\eqref{eq:lemma_convolution_proof} and calculating the integral, we obtain Eq.~\eqref{eq:lemma_convolution}. Notice that the $\mathbf{P}(x)$ in the RHS of Eq.~\eqref{eq:lemma_convolution} may be higher in  degree~(higher by at most 1) than the $\mathbf{P}(x)$ in the LHS, since the summation $\sum^{R_1}_{r=1}r^{-\gamma}$ introduces an additional  $\ln R$ factor when $\gamma=1$. 
\subsection{Derivation of the bounds in higher dimension}\label{app:FSE_higherD}
\begin{figure}
	\center{\includegraphics[width=0.5\linewidth]{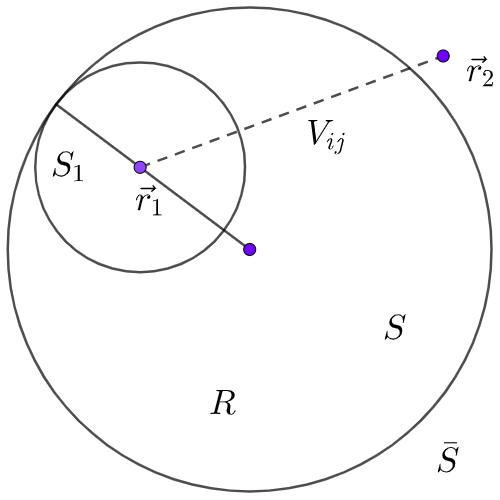}}
	\caption{\label{fig:FSE-higherD} Derivation of the finite-size error bound in higher spatial dimension. $S$ is the finite system with radius $R$, $\overline{S}$ is its complement, $\vec{r}_1, \vec{r}_2$ are respectively the positions $i,j$ of the power-law interaction $V_{ij}$, $S_1\subseteq S$ is a subsystem centered at $\vec{r}_1$ with radius $R-r_1$~(so that $S_1$ touches $S$).
	}
\end{figure}

In Sec.~\ref{sec:FSE} we derived the finite size error bound Eq.~\eqref{eq:FSEB} in 1D. In the following we generalize the derivation to arbitrary spatial dimension. The configuration is shown in Fig.~\ref{fig:FSE-higherD}. Without loss of generality we can assume that the system has a spherical shape~(the sphere $S$ in Fig.~\ref{fig:FSE-higherD}), since the error in other cluster shapes can be upper and lower bounded by spheres with radius proportional to its linear dimension.  Eq.~\eqref{eq:FSE_sum} is still valid, so we have
\begin{eqnarray}\label{eq:FSE_higherD1}
	|\delta\langle \hat{S}_X \rangle_{L}| &\leq&  \sum_{\vec{r}_1\in S,\vec{r}_2\in \overline{S}} \frac{\mathbf{P}(\ln r_1)}{r_1^{\alpha_1}|\vec{r}_2-\vec{r}_1|^{\alpha}}\nonumber\\
	 &\leq&  \sum_{\vec{r}_1\in S,\vec{r}_2\in \overline{S_1}} \frac{\mathbf{P}(\ln r_1)}{r_1^{\alpha_1}|\vec{r}_2-\vec{r}_1|^{\alpha}}\nonumber\\
	 &\leq&  \sum_{\vec{r}_1\in S} \frac{\mathbf{P}(\ln r_1)}{r_1^{\alpha_1}(R-r_1)^{\alpha-D}}\nonumber\\
	 &\leq&  \sum_{r_1=1}^{R-1} \frac{\mathbf{P}(\ln r_1)}{r_1^{\alpha_1-D+1}(R-r_1)^{\alpha-D}},
\end{eqnarray}
where in the second line we extended the sum in $\vec{r}_2$ from $\overline{S}$ to $\overline{S_1}$~(which contains $\overline{S}$), in the third and the last lines we upper bounded the sums by integration~[with a constant coefficient absorbed into $\mathbf{P}(\ln r_1)$], and the integrals can be calculated analytically due to the spherical geometry.
Applying Lemma~\ref{lemma:power_law_convolution} to Eq.~\eqref{eq:FSE_higherD1}, we obtain Eq.~\eqref{eq:FSEB} with $\alpha_3$ summarized in Tab.~\ref{tab:comp}. 

\section{Bounds for correlation decay: proof that $|\Omega_{XY}(i y)|\leq |\bar{\Omega}_{XY}(0)|$}\label{app:MaxOmega}
In this section we prove the claim we made in Sec.~\ref{sec:correlation} that $|\Omega_{XY}( i y)|$ for $y\in\mathbb{R}$ can always be upper bounded by the upper bound of $|\Omega_{XY}(0)|$ obtained by Eq.~\eqref{eq:holographicbound}. We prove this for the $\Omega_{XY}(\omega)$ in Sec.~\ref{sec:LPPL_small_alpha}, i.e. power-law systems with $\alpha>D$, and the proofs for other cases are similar.  

We begin by recalling a simple fact about subharmonic functions: if $p(\omega)$ is a real-valued subharmonic function and $q(\omega)$ is a real-valued harmonic function such that $p(\omega)\leq q(\omega)$ on the boundary of a simply-connected domain $S$, then $p(\omega)\leq q(\omega)$ everywhere in $S$. Now take $p(\omega)$ to be the subharmonic function $\ln |\Omega_{XY}(\omega)|$ and take $q(\omega)$ to be the unique harmonic function which agrees with $\ln \bar{\Omega}(r,y)$ on the boundary of the region $S_\rho$ bounded by the parametric curve $f(\rho e^{i\theta}), \theta\in[0,2\pi]$, as plotted in Fig.~\ref{fig:omega_plane}, where $f(z)$ is defined in Eq.~\eqref{eq:g_v_conformal}, $\rho\in(0,1)$, and later we will consider the limit $\rho\to 1$. By construction, we have $p(\omega)\leq q(\omega)$ on $\partial S_\rho$, therefore $p(\omega)\leq q(\omega)$ everywhere in $S_\rho$. Using the mean-value property of harmonic functions, in the limit $\rho\to 1$ we have 
\begin{eqnarray}
	\lim_{\rho\to 1}q(0)&=&\lim_{\rho\to 1}\frac{1}{2\pi}\int_0^{2\pi}q[f(\rho e^{i\theta})]d\theta\nonumber\\
	&=&\lim_{\rho\to 1}\frac{1}{2\pi}\int_0^{2\pi}\ln \bar{\Omega}[r,|\mathrm{Im}f(\rho e^{i\theta})|] d\theta\nonumber\\
	&=&\ln |\bar{\Omega}_{XY}(0)|.
	\end{eqnarray}
Therefore, to prove that $|\Omega_{XY}(i y)|\leq |\bar{\Omega}_{XY}(0)|$, it suffices to prove that $q(i y)$ is monotonically decreasing in $y$ for $y\geq 0$. In the following we prove this for any $\rho\in (0,1)$. 

Since $q(\omega)$ is harmonic, for illustrative purpose we use the language of electrostatics. From the expression of $ \bar{\Omega}(r,y)$ in  Eq.~\eqref{eq:Gry_smallalpha} it is clear that on the boundary of $S_\rho$ the potential $q(\omega)$ is strictly decreasing in the direction of increasing $|y|$. In the following we use proof by contradiction: if $q(i y)$ is not monotonically in $y$ for $y\geq 0$, then there must exist $y_1,y_2$ satisfying $0<y_1<y_2<f(i\rho)/i$ such that $q(i y_1)=q(i y_2)$. Let $l_1,l_2$ be the equipotential lines passing through $iy_1,iy_2$, respectively. Equipotential lines cannot terminate in free space, since otherwise it would imply there is an electric charge at the end point. $l_1$ and $l_2$ cannot intersect anywhere, since for example if they intersect at a point $x+iy$ with $x>0,y>0$, then by symmetry they also intersect at $-x+iy$, which implies that $l_1,l_2$ enclose a region in which $q(\omega)$ is a constant, which is impossible for a non-constant harmonic function. By similar logic~(and using the mirror symmetry with respect to the real axis) neither $l_1$ nor $l_2$ can intersect with the real axis, so we can focus our attention on the upper half plane. Furthermore, at most one of $l_1,l_2$ can intersect with the boundary of $S_\rho$, since  $q(\omega)$ is strictly decreasing in the direction of increasing $|y|$ on the boundary. Without loss of generality suppose $l_1$ does not intersect the boundary. Then the only remaining possibility is that $l_1$ is a closed curve inside $S_\rho$. But this implies that $q(\omega)$ is constant in the interior of $l_1$, which is impossible for a non-constant harmonic function. %
In conclusion, $q(iy)$ must be monotonically decreasing in $y$ for $y>0$. [It is straightforward to rule out the possibility of $q(iy)$ being monotonically increasing in $y$: since if that's the case there must exist $x,y$ with $0<x<f(\rho),0<y<f(i\rho)/i$ such that $q(x)=q(iy)$. Considering the equipotential curve passing through $x$, $iy$, $-x$, and $-iy$, we reach a similar contradiction.]

\bibliography{/home/lagrenge/Documents/Mendeley_bib/library,LPPLfootnotes}

\end{document}